\RequirePackage{fix-cm}
\documentclass[smallextended]{svjour3}       
\smartqed  
\usepackage{graphicx}
\usepackage[misc]{ifsym}
\usepackage{xcolor}
\usepackage{lineno}

\newcommand{\absdiv}[1]{%
		\par\addvspace{.5\baselineskip}
	\noindent\textbf{#1}\quad\ignorespaces
}
\begin{document}

\title{Characterization of the first prototype CMOS pixel sensor developed for the CEPC vertex detector
}



\author{L.J. Chen\textsuperscript{1,2,3}  \and
	H.B. Zhu\textsuperscript{1,3}  \and
	X.C. Ai\textsuperscript{1,4}  \and
	M. Fu\textsuperscript{6}  \and
	R. Kiuchi\textsuperscript{1,3}  \and
	Y. Liu\textsuperscript{4}  \and
	Z.A. Liu\textsuperscript{1,3}  \and
	X.C. Lou\textsuperscript{1,3}  \and
	Y.P. Lu\textsuperscript{1,3}  \and
	Q. Ouyang\textsuperscript{1,3}  \and
	X. Shi\textsuperscript{1,3}  \and
	J. Tao\textsuperscript{1,2,3}  \and
	K. Wang\textsuperscript{1,3}  \and
	N. Wang\textsuperscript{1,3}  \and
	C.F. Yang\textsuperscript{3,5}  \and
	Y. Zhang\textsuperscript{1,3}  \and
	Y. Zhou\textsuperscript{1,3} 
}

\institute{\Letter  ~H.B. Zhu\\
	\email{zhuhb@ihep.ac.cn}
	\\
	\\
	\textsuperscript{1} Institute of High Energy Physics, Chinese Academy of Sciences, 19B Yuquan Road, Shijingshan District, Beijing, China \\
	\textsuperscript{2} University of Chinese Academy of Sciences, No.19(A) Yuquan Road, Shijingshan District, Beijing, China\\
	\textsuperscript{3} State Key Laboratory of Particle Detection and Electronics, 19B Yuquan Road, Shijingshan District, Beijing, China\\
	\textsuperscript{4} Deutsches Elektronen-Synchrotron (DESY),  Notkestra$\beta$e 85 D-22607 Hamburg, Germany\\
	\textsuperscript{5} University of Science and Technology of China, No.96, JinZhai Road Baohe District, Hefei, Anhui, China\\
	\textsuperscript{6} Ocean University of China,  No. 238 Songling Road, Qingdao 266100, China
}


\date{Received: date / Accepted: date}

\maketitle

\begin{abstract}

\hspace{0pt}
\absdiv{Purpose} CMOS pixel sensors have become extremely attractive for future high performance tracking devices. Initial R\&D work has been conducted for the vertex detector for the proposed Circular Electron Positron Collider that will allow precision Higgs measurements. It is critical to achieve low power consumption to minimize the material budget. This requires careful optimization of the sensor diode geometry to reach high charge-over-capacitance that allows reduction in analog power consumption.

\absdiv{Methods}
The electrode area and footprint are two critical elements in sensor diode geometry and have deciding impacts on the sensor charge collection performance. Prototype CMOS pixel sensor JadePix-1 has been developed with pixel sectors implementing different electrode area and footprint and their charge collection performance has been characterized with radioactive resources. 

\absdiv{Results}
Charge-to-voltage conversion gains are calibrated with low energy X-ray. Noise, charge collection efficiency, charge-over-capacitance and signal-to-noise ratio are obtained for pixel sectors of different electrode area and footprint.

\absdiv{Conclusion}
Small electrode area and large footprint are preferred to achieve high charge-over-capacitance that promises low analog power consumption. Ongoing studies on sensor performance before and after irradiation, combined with this work, will conclude on the diode geometry optimization. 


\end{abstract}

\section{Introduction}
\label{introduction}
The proposed Circular Electron Positron Collider (CEPC)~\cite{CEPC_Acc_CDR,CEPC_Detector_CDR} will function as a Higgs factory and allow measurements of the Higgs properties with high precision beyond the Large Hadron Collider (LHC) and its successor, the High Luminosity LHC (HL-LHC)~\cite{HL-LHC_ATL_Higgs,HL-LHC_CMS_Higgs}. To achieve the required tracking precision that is essential for heavy-flavor tagging and $\tau$-tagging, the CEPC vertex detector will have to be built with the state-of-the-art pixel detector technologies. It is desirable to achieve spatial resolution better than 3~$\mu$m, power consumption below 50~mW/cm$^2$ and material budget below 0.15\%$X_0$ for each detector layer. In addition, the vertex detector must sustain with radiation damage of about 1~MRad/year (Total Ionizing Dose) and $2\times10^{12}~\mathrm{1~MeV}~n_{eq}/\mathrm{cm}^2$/year (Non-Ionizing Energy Loss) during the detector operation at the center-of-mass energy of $\sqrt{s}=240$~GeV.

Among different pixel detector technologies, CMOS pixel sensors represent the most promising one for the CEPC vertex detector. They allow the integration of the active detection element and its readout electronics on the same silicon substrate. Advantageous features, including small pixel size, low power consumption and low material budget, make them extremely attractive for charged-particle tracking~\cite{MAPS-1,MAPS-2,MAPS-3} in high energy physics. Such pixel sensors were successfully deployed for the high precision EUDET beam telescopes~\cite{EUDET} and the STAR-HFT at RICH~\cite{STAR_HFT}. They are already selected for the ALICE ITS Upgrade~\cite{ALICE_ITS_TDR,ALICE_ITS_ALPIDE} and being actively explored for the ATLAS Phase II Inner Tracker Pixel Upgrade~\cite{ATLAS_ITK_PIX_TDR,ATLAS_ITK_PIX_TJ,ATLAS_ITK_PIX_AMS,ATLAS_ITK_STRIP}. 

In CMOS pixel sensor designs, it is of primary interest to achieve sufficient signal charge-over-capacitance ($Q/C$), which can lead to low analog power consumption that is crucial for reducing material budget. As described in~\cite{MAPS_QC}, the analog power consumption depends on the signal-to-noise ($S/N$) ratio required for a given bandwidth. Since the leakage current noise can be reduced by decreasing the integration time, the thermal noise of the input transistor becomes the dominant source. The power consumption is dominated by the bias current ($I_{bias}$) of the input transistor and strongly depends on the $Q/C$ ratio as follows:

\begin{equation}
P_{analog} \propto I_{bias} \propto \left(\frac{S/N}{Q/C}\right)^{2m}
\end{equation}
\noindent where $S$ represents the signal charge, $N$ the noise, $Q$ the collected charge, and $C$ the input capacitance. Parameter $m$ depends on the transistor operating point and takes $m=2$ in strong inversion and $m = 1$ in weak inversion. For a given $S/N$ and bandwidth, a higher $Q/C$ allows for lower power consumption. In this paper, detailed designs of the first prototype CMOS pixel sensor JadePix-1 for the CEPC vertex detector are described and the test results for different diode geometries to achieve high $Q/C$ are presented.

\section{Sensor Design}
\label{sec:pixel_design}

\begin{figure}[!htp]
	\centering
	\includegraphics[width=0.80\textwidth]{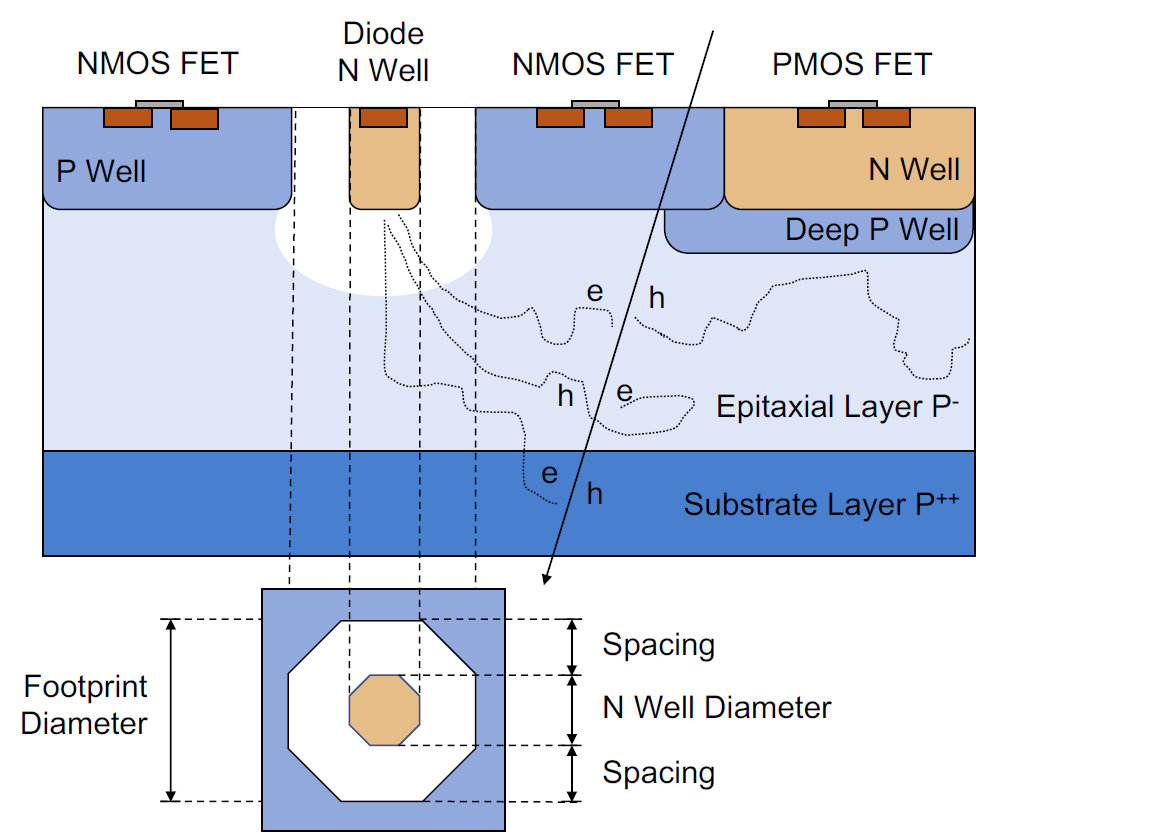}
	\caption{Schematic cross section of JadePix-1 fabricated with a novel 0.18~$\mu$m CMOS imaging sensor process with the shielding deep P-well feature. A small depletion zone is formed around the N-well diode.}
	\label{fig:JadePix_XS}       
\end{figure} 

With a 0.18~$\mu$m CMOS imaging sensor process, the first prototype sensor JadePix-1 for the CEPC vertex has been developed to evaluate the charge collection performance and optimize the sensor diode geometry to achieve a high $Q/C$ ratio~\cite{JadePix_Design}. As illustrated in Figure~\ref{fig:JadePix_XS}, the collection diode is formed by an octagon shape N-implant electrode in the P-doped epitaxial layer, which extends for 18~$\mu$m in depth and features high resistivity of $\rho > 1~\mathrm{k}\Omega\cdot$cm. The novel CMOS process offers a deep P-well implantation underneath the N-well that houses the PMOS transistors. The deep P-well provides effective shielding and prevent deposited signal charge distracted by the non-electrode N-well. This allows for circuit design with both NMOS and PMOS transistors and enhances the in-pixel readout electronics functionalities. In this pixel sensor structure, only a small depletion zone can be formed around the N-well electrode. Due to the lack of electric field in the active detector volume, charge collection is predominantly achieved through thermal diffusion. The charge carriers diffuse and continuously get reflected between the potential barriers formed in the interfaces of the epitaxial layer and the substrate, as well as of the epitaxial layer and the P-well until they get collected at the collection electrode. Each pixel diode consists of an octagonal shape N-well electrode surrounded by the P-well. Additional space between the two is introduced to mitigate the potential drop of charge collection efficiency caused by the reduced N-well area. The footprint represents the total area confined by the surrounding P-well. Both electrode area and footprint play critical roles in charge collection and their impacts on charge collection performance must be carefully evaluated.

\begin{figure}[!htp]
	\centering
	\includegraphics[width=0.8\textwidth]{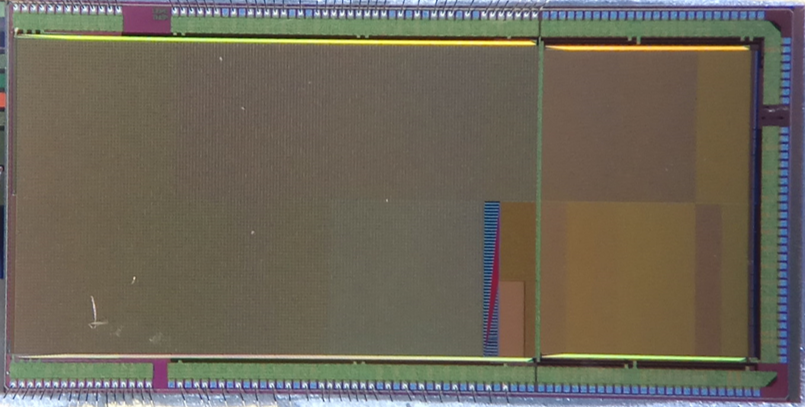}
	\caption{The JadePix-1 prototype sensor segmented into 20 sectors on the left side and 16 sectors on the right side, respectively.}
	\label{fig:JadePix}       
\end{figure}

The prototype sensor JadePix-1, as shown in Figure~\ref{fig:JadePix}, consists of two matrices. The left matrix contains 19 sectors of $33\times33~\mu\mathrm{m}^{2}$ pixels implemented with different diode variants or test structures, and one sector of $16\times16~\mu\mathrm{m}^{2}$ pixels with the diode that can be biased with an external voltage. Each sector implements a collection electrode with different sizes of N-well and footprint. Pixels are arranged in 48 columns and 16 rows. The right matrix contains 16 sectors of $16\times16~\mu\mathrm{m}^{2}$ pixels that are arranged in 96 columns and 16 rows in each sector. Conventional 2/3~T readout structures are implemented to read out the charge signals. The pixel sector under investigation can be selected and read out individually in the rolling-shutter mode at the clock frequency of 2~MHz. In the following, test results obtained for sectors in the upper row of the left matrix with $33\times33~\mu\mathrm{m}^{2}$ pixels are presented and the impacts of diode geometries are evaluated.


%


\section{Data Acquisition and Processing}
\label{sec:daq_process}
\begin{figure}[!htp]
	\centering
	\includegraphics[width=0.8\textwidth]{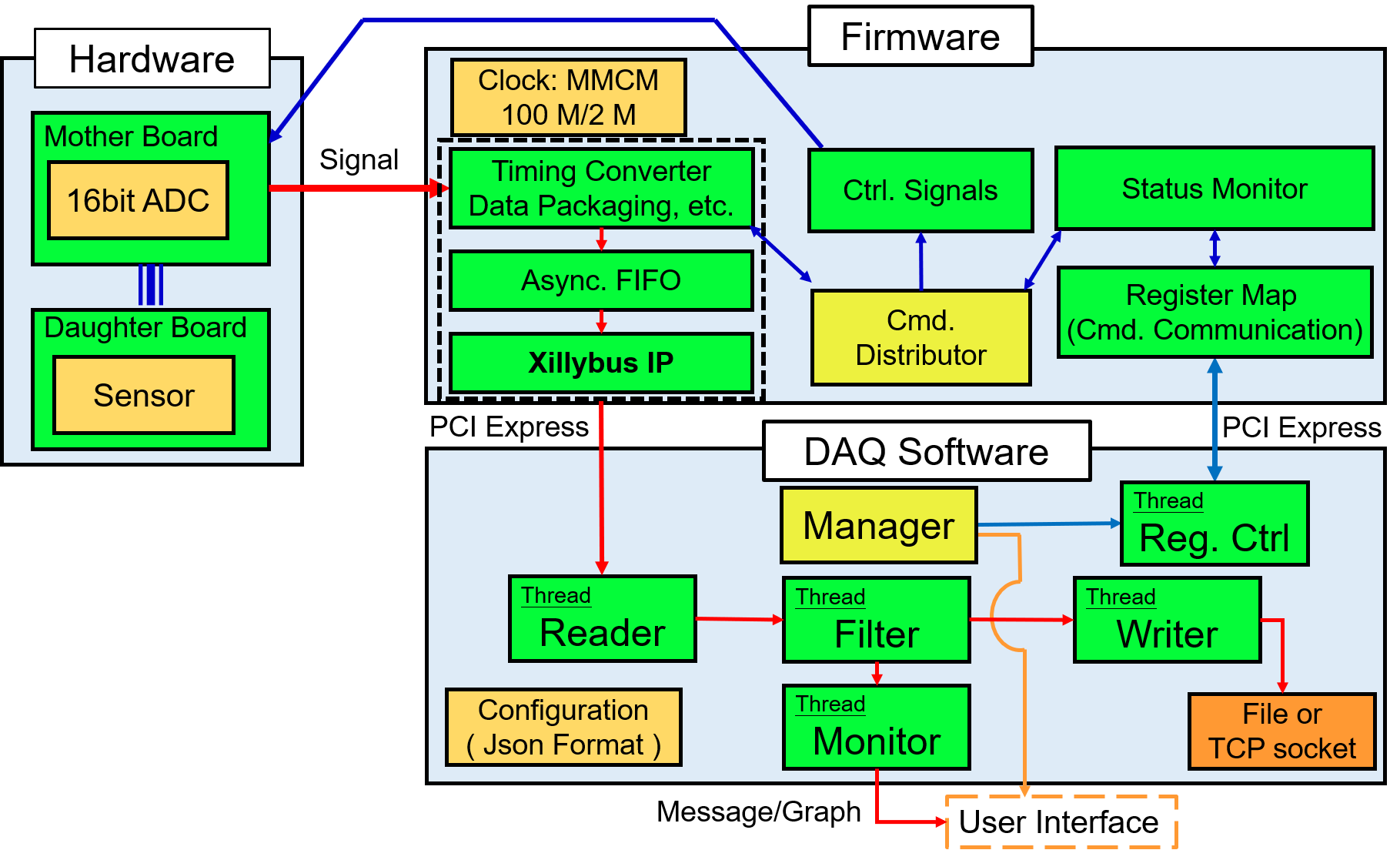}
	\caption{Structure of the data acquisition system developed for JadePix-1 testing.}
	\label{fig:JadePix_DAQ}       
\end{figure}

The data acquisition (DAQ) system, as sketched in Figure~\ref{fig:JadePix_DAQ}, is conceived to easily characterize the prototype JadePix-1 sensor, providing low noise and high speed. The DAQ hardware consists of daughter board, mother board, FPGA board and PC, and supports up to 16 readout channels. The daughter board amplifies the single-ended signal from the JadePix-1 sensor and converts it to a differential signal, which allows long-distance data transmission. The mother board is designed to receive and digitize the analogue signal from the daughter board using multiple high precision 16-bit analog-to-digital converters (ADCs). In addition, it provides powering for the daughter board and the sensor under test. The system noise is measured to be below 3.5~$e^{-}$ equivalent noise charge (ENC) at the input of the amplifier and sufficiently low for sensor characterization. A commercial FGPA evaluation board (KC705) is deployed to read in the digitized signals from the mother board and assemble them into event data frames before sending them to the control PC via PCIe. It also sends out clocks and control commands for data taking. The system supports high speed data transmission up to 6Gbps. 

To facilitate efficient data taking, a multi-threading DAQ software has been developed based on the combination of C++/python. Collected data are assembled into event frames and stored on large volume data disks for off-line data processing. The software also supports on-line monitoring with sampled events. During the off-line data processing, the correlated double sampling (CDS) signal processing technique has been used to extract pixel signals from the difference between two consecutive frames. CDS acts similarly to a high-pass filter for noise, suppressing low-frequency ($1/f$) noise and mitigating offset in slow-changing analog signals, which are critical for the CMOS pixel sensor performance.

\section{Test Results}
\label{sec:results}

\subsection{Charge-to-Voltage Conversion Gain Calibration}

\begin{figure}[!htp]
	\centering
	\includegraphics[width=0.8\textwidth]{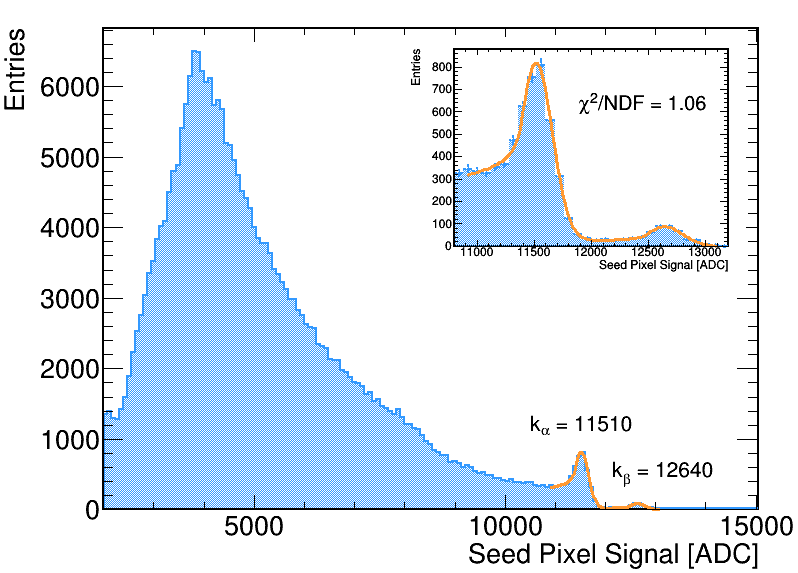}
	\caption{Signal distribution for photons emitted by the $^{55}$Fe radioactive source measured with the seed pixel, which contains the highest charge within a $3\times3$ pixel cluster.}
	\label{fig:JadePix_Seed}       
\end{figure}

The collected charge in a CMOS pixel sensor can be sensed as a voltage variation on the diode. The charge-to-voltage conversion gain is defined as the ratio of the signal voltage after the source follower to the collected charge carriers, \textit{i.e.} electrons during the signal integration time. This basic characteristic parameter can be estimated with the emission spectra of a low energy X-ray radioactive source, e.g. $^{55}$Fe. The emitted 5.9~keV X-rays (a.k.a. $k_{\alpha}$) generate constantly about 1640 $e-h$ pairs in the active sensor volume. As described above, most of the charges are collected by the collection electrode through the slow thermal diffusion process, with a collection time of about 100~ns. This represents the first peak at low signal amplitude in the $^{55}$Fe spectra as shown in Figure~\ref{fig:JadePix_Seed}. However, for photons that are converted near the collection electrode, the generated $e-h$ pairs are separated effectively and charges can be quickly collected by the electrode due to the presence of high electric field around. This leads to a charge collection efficiency close to 100\% and represents the second peak at high amplitude in the $^{55}$Fe spectra in Figure~\ref{fig:JadePix_Seed}. The formed $k_{\alpha}$ peak can be used to calibrate the charge-to-voltage conversion gain. A crystal ball function has been adopted to fit the spectra and to extract the exact $k_{\alpha}$ peak position. It describes better the incomplete charge collection at the edge of the small depletion zone, resulting in a non-Gaussian tail on the lower side of the $k_{\alpha}$ peak. Thanks to the excellent performance of JadePix-1, it is possible to observe the third but lower peak formed by the complete collection of charges generated by the impinging $k_{\beta}$ X-rays of 6.4~keV. Following the calibration procedure defined in~\cite{MAPS_CALIB}, the charge-to-voltage conversion gain can be defined as:

\begin{equation}
\mathrm{Conversion~Gain}~[\mu V/e]= \frac{V_{\mathrm{calib}}\times 10^{6}~ [\mu V/V]}{1640~e^{-}}
\end{equation}

\noindent where 1640 $e^{-}$ is the total charges generated by a 5.9 keV photon. $V_{\mathrm{calib}}$ is the calibrated voltage after the source follower and determined by:

\begin{equation}
V_{\mathrm{calib}} = k_{\alpha} ~[\mathrm{ADC}] \times \frac{2.5~[V]}{2^{N}-1} \times\frac{1}{g_a} 
\end{equation} 

\noindent where $k_{\alpha}$ is the ADC counts of the $k_{\alpha}$ peak position in the $^{55}$Fe spectra, $N=16$ the total bits of the ADC, and $g_a=8$ the gain at the stage of differential amplification.

\begin{figure}[!htp]
	\centering
	\includegraphics[width=0.49\textwidth]{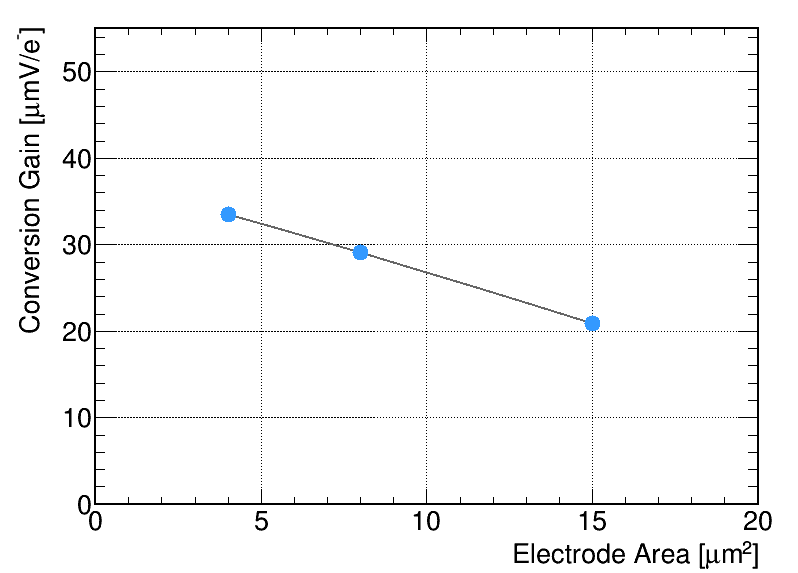}
	\includegraphics[width=0.49\textwidth]{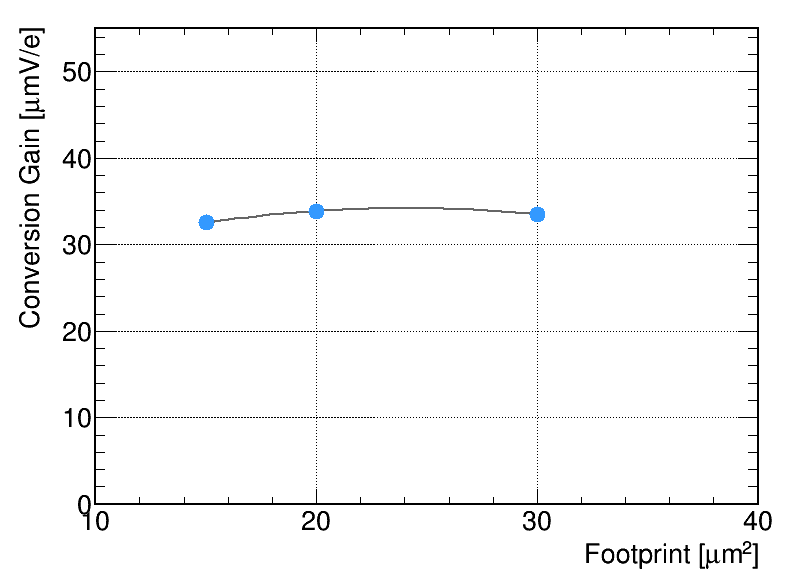}
	\caption{The measured charge-to-voltage conversion gain decreases with higher electrode area but remains nearly the same for different footprint.}
	\label{fig:JadePix_Gain}       
\end{figure}

The charge-to-voltage conversion gains are evaluated for each pixel sector. The statistical uncertainties on conversion gains are estimated to be around 1\% and the gain variations across each sector are found to be below 0.5\%. Figure~\ref{fig:JadePix_Gain} shows that the charge-to-voltage conversion gain decreases as the electrode area increases from 4~$\mu \mathrm{m}^{2}$ to 15~$\mu \mathrm{m}^{2}$, but reveals nearly no dependence on the footprint varying from 15~$\mu \mathrm{m}^{2}$ to 30~$\mu \mathrm{m}^{2}$. This can be explained by that the larger electrode introduces higher diode capacitance, which leads to lower gain, but the footprint has only modest impacts on the diode capacitance.

\begin{figure}[!htp]
	\centering
	\includegraphics[width=0.49\textwidth]{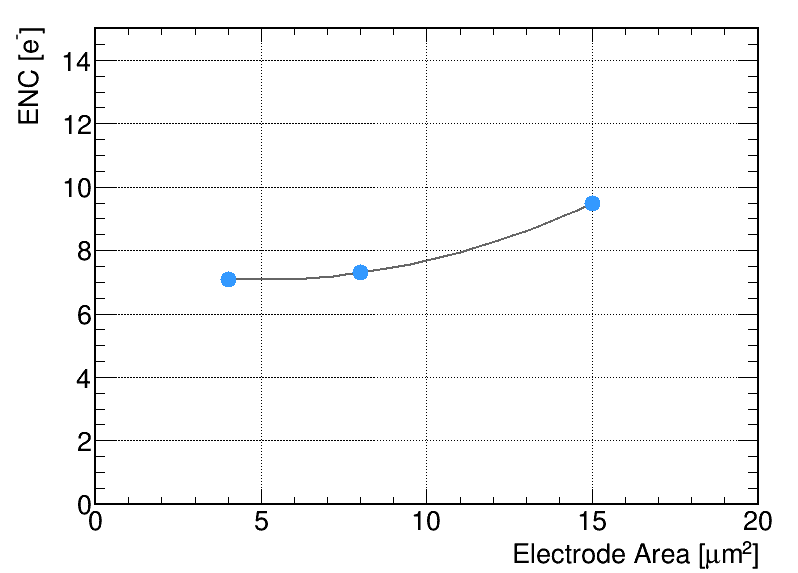}
	\includegraphics[width=0.49\textwidth]{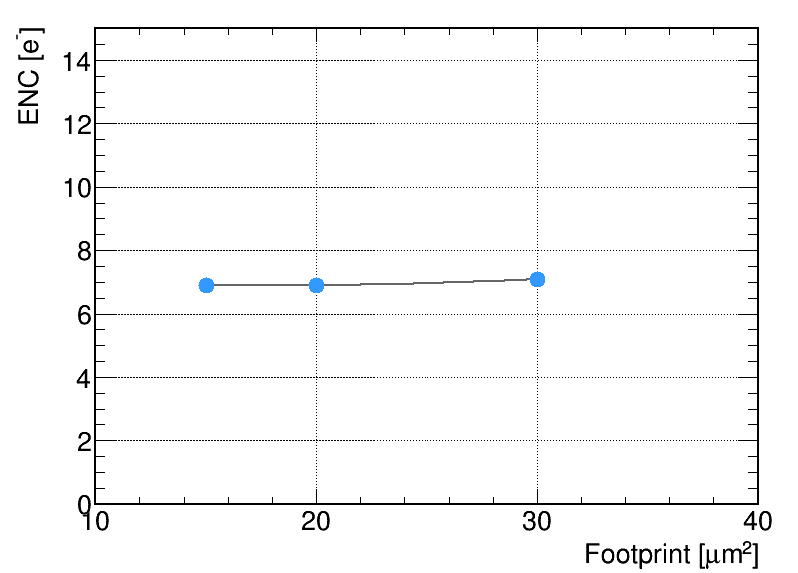}
	\caption{The measured ENC noise increases with larger electrode area but stays almost the same for increasing footprint.}
	\label{fig:JadePix_Noise}       
\end{figure}

\subsection{Noise}

In pixel sensor design, it is always desirable to reduce the noise and increase the signal-to-noise ratio. To measure the noise after the CDS signal processing, the JadePix-1 sensor needs to be placed in a light-shielded box and protected from any external signal. The results are then converted to equivelent noise charge (ENC) using the charge-to-voltage conversion gain derived for each pixel sector. Due to the presence of random telegraph signal (RTS) noise~\cite{RTS} that leads to non-Gaussian tails in the noise distributions, the RMS values are taken as the noise for each sector. Figure~\ref{fig:JadePix_Noise} shows that noise increases for larger electrode area but does not change with varying footprint.



\subsection{Charge Collection Efficiency}
\begin{figure}[!htp]
	\centering
	\includegraphics[width=0.49\textwidth]{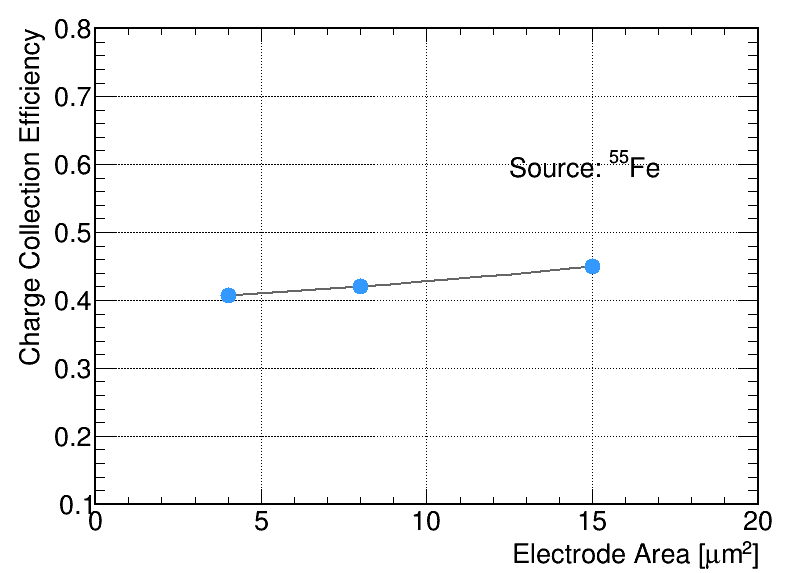}
	\includegraphics[width=0.49\textwidth]{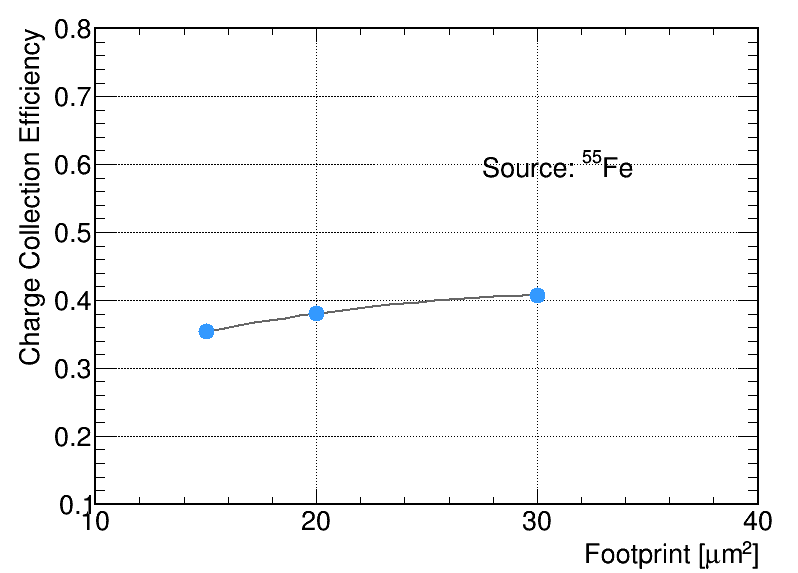}
	\caption{The measured charge collection efficiency with the $^{55}$Fe radioactive source. The CCE is defined as the most probable value of charge collected by the seed pixel via thermal diffusion divided by the full charge, \textit{i.e.} the $k_{\alpha}$ peak position.}
	\label{fig:JadePix_CCE}       
\end{figure}

The charge collection efficiency represents the sensor capability of collecting the charges generated by the impinging particles. As for the tests performed with a $^{55}$Fe radioactive source, the charge collection efficiency is simply defined as the charge collected by the seed pixel via thermal diffusion (the collection peak) divided by the full deposited charge, or equivalently the $k_{\alpha}$ peak position. Figure~\ref{fig:JadePix_CCE} shows the increasing charge collection efficiency for larger electrode area and larger footprint. In both cases, the expanded depletion zone shortens the collection time of charge carriers and improves the charge collection efficiency. 

\begin{figure}[!htp]
	\centering
	\includegraphics[width=0.49\textwidth]{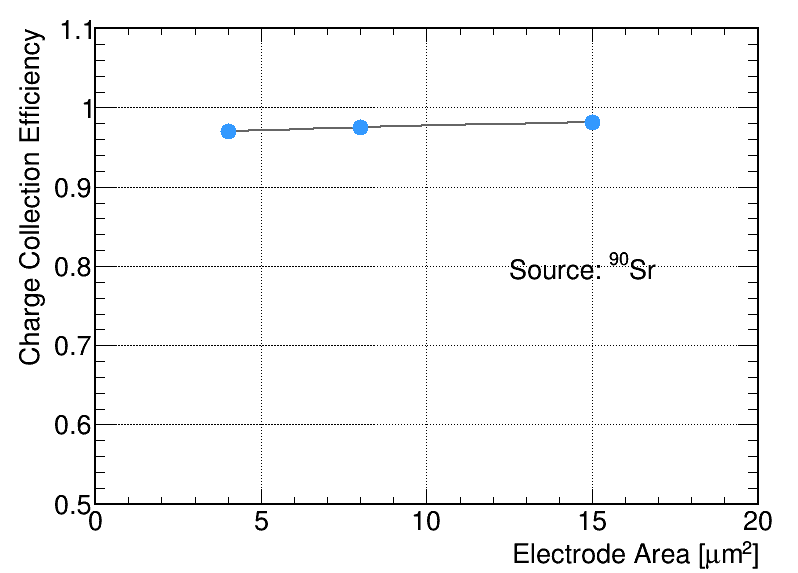}
	\includegraphics[width=0.49\textwidth]{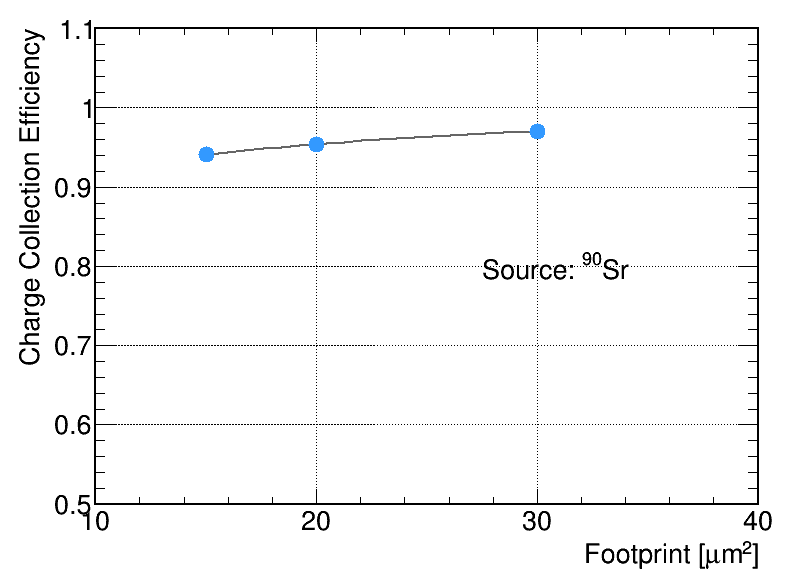}
	\caption{The measured charge collection efficiency with a $^{90}$Sr radioactive source. The charge collection efficiency is re-defined as the charges collected in $3\times 3$ clusters divided by the charges collected in $5\times5$ clusters}
	\label{fig:JadePix_CCE_Sr}       
\end{figure}

The charge collection efficiency has been also characterized with a $^{90}$Sr radioactive source. The total deposited charges, following a Landau distribution, are in general unknown but studies show that the generated charges are almost fully contained with a sufficiently large pixel cluster, e.g. $5\times5$ cluster. Therefore the charge collection efficiency is re-defined as the most probable value (MPV) derived from the Landau fit to the $3\times3$ cluster charge distribution divided by the one derived from the fit to the $5\times5$ cluster charge distribution. Similarly, as shown in Figure~\ref{fig:JadePix_CCE_Sr}, higher charge collection efficiencies are achieved with larger electrode area and larger footprint.


\subsection{Charge-over-Capacitance Ratio}

\begin{figure}[!htp]
	\centering
	\includegraphics[width=0.49\textwidth]{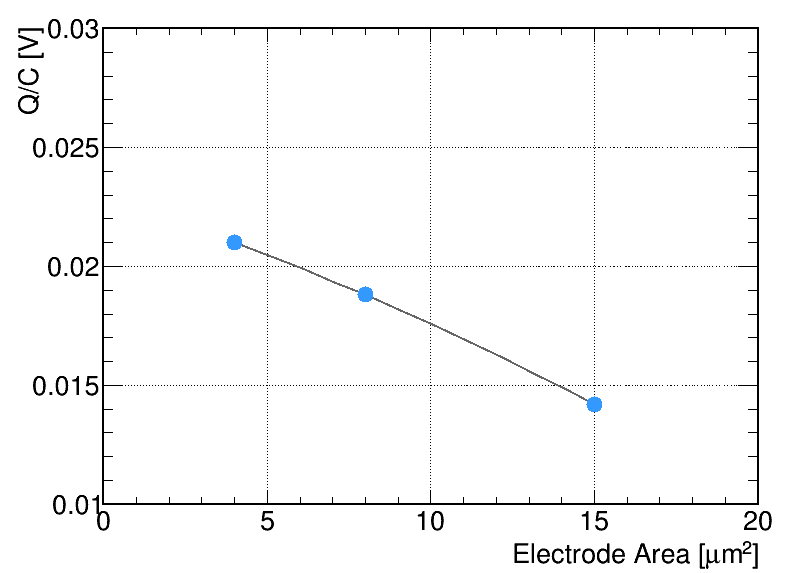}
	\includegraphics[width=0.49\textwidth]{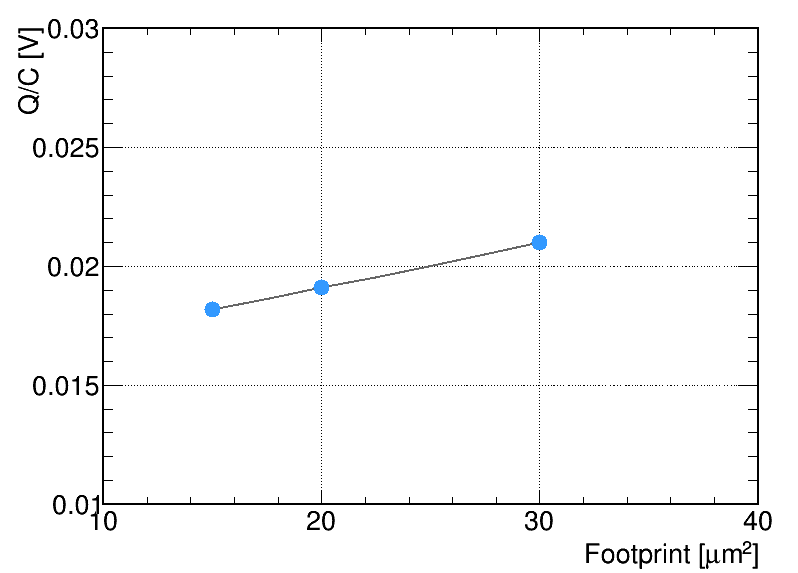}
	\caption{The measured Q/C decreases with larger electrode area and increases with larger footprint.}
	\label{fig:JadePix_QC}       
\end{figure}

As mentioned before, high $Q/C$ ratio plays a key role in reducing the analog power consumption and requires proper optimization of the diode geometry. While the effective charge $Q$ is taken from the MPV of the collected charges in seed pixels in response to the $^{90}$Sr radioactive source, the input capacitance can be derived as follows:

\begin{equation}
C~[fF] = \frac{1640~[e] \times 1.6 \times 10^{-19}~[C/e]}{V_{\mathrm{calib}}/g_b} \times 10^{15}~[F/fF]
\end{equation}

\noindent where $g_b = 0.8$ represents the gain of the source follower in the 2T readout structure. Figure~\ref{fig:JadePix_QC} shows that small electrode and large footprint can yield high $Q/C$ that can lead to low analog power consumption. This reveals the fact that increasing electrode area and the spacing between the collection N-well to surrounding P-well allows for wider depletion volume and lowers the input capacitance. Although the charge collection efficiency, and equivalently the effective charge, drops for the small electrode, it still yields high $Q/C$ and large footprint is always preferred as far as it is allowed by the pixel size.

\begin{figure}[!htp]
	\centering
	\includegraphics[width=0.49\textwidth]{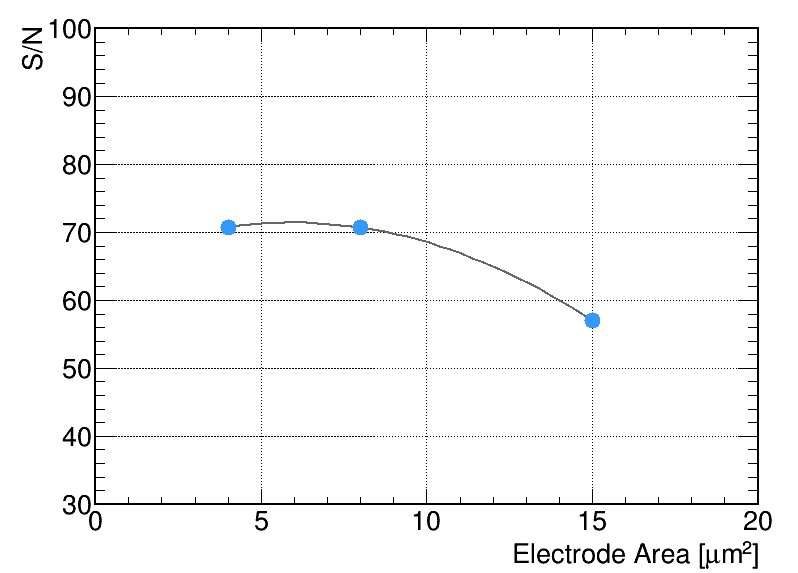}
	\includegraphics[width=0.49\textwidth]{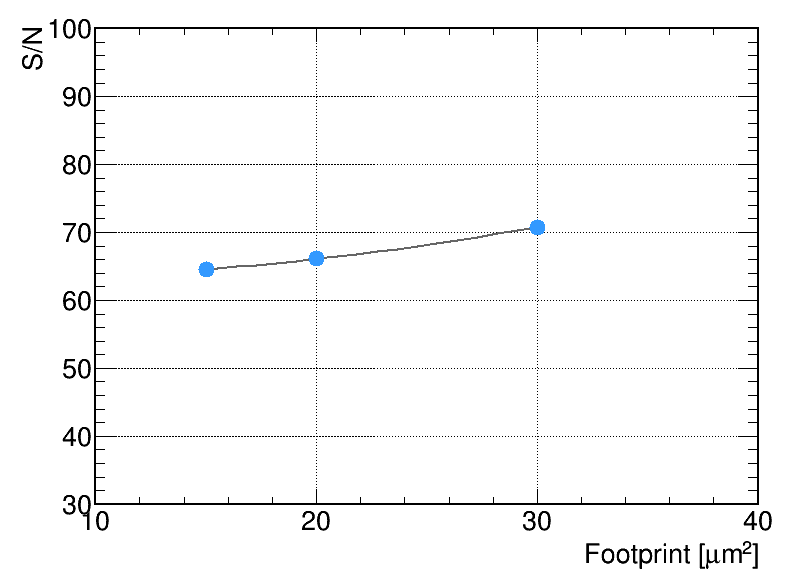}
	\caption{The measured signal-to-noise ratio decreases with larger electrode area but increases with larger footprint.}
	\label{fig:JadePix_SNR}       
\end{figure}

In addition, the important signal-to-noise ratios are calculated. While the signal is taken as the same effective charge $Q$ described above, the noise is determined from the ENC noise for the corresponding pixel sector as derived before. As shown in Figure~\ref{fig:JadePix_SNR}, small electrode and large footprint result in high $S/N$ that is always preferred for efficient detector operation.
 
\section{Conclusion}
The first prototype CMOS pixel sensor JadePix-1 has been developed for the CEPC vertex detector. Its sensor designs have been described in detail. The noise and charge collection efficiency have been validated with radioactive sources for each sector. The results presented show that small electrode area and large footprint are preferred to achieve high $Q/C$ that is critical to reduce the analog power consumption as required for the CEPC vertex. More studies are being carried out to evaluate the charge collection performance before and after irradiation that combined with the $Q/C$ performance will conclude on the the diode geometry optimization.


\begin{acknowledgements}
This project is jointly supported by the National Natural Science Foundation of China (No. 11505207), the State Key Laboratory of Particle Detection and Electronics, the CAS Center for Excellence in Particle Physics (CCEPP), the IHEP Innovation Fund and the International Partnership Program of Chinese Academy of Sciences. 
\end{acknowledgements}



\end{document}